\RequirePackage{fix-cm}

\documentclass{svjour3}                     

\smartqed  

\usepackage{graphicx}
\usepackage{dcolumn}
\usepackage{braket,amsmath,amssymb,bm}
\newcommand{\NOON}{N00N }

\begin{document}

\title{Generation of phase-squeezed optical pulses with large coherent amplitudes by post-selection of single photon and weak cross-Kerr non-linearity}

\titlerunning{Generation of phase-squeezed optical pulses with large coherent amplitudes}

\author{Fumiaki Matsuoka         \and
        Akihisa Tomita         \and
        Yutaka Shikano
}


\institute{F. Matsuoka \and A. Tomita \at 
              Graduate School of Information Science and Technology, Hokkaido University, Kita 14 - Nishi 9, Kita-ku, Sapporo, 060-0814, Japan \\
              \email{matsuoka@optnet.ist.hookudai.ac.jp}             
           \and
           F. Matsuoka \and Y. Shikano \at
              Research Center of Integrative Molecular Systems (CIMoS), Institute for Molecular Science, National Institutes of Natural Sciences, 38 Nishigo-Naka, Myodaiji, Okazaki, Aichi, 444-8585, Japan           
           \and
           Y. Shikano \at
              Institute for Quantum Studies, Chapman University, 1 University Dr., Orange, CA 92866, USA\\
              Materials and Structures Laboratory, Tokyo Institute of Technology, 4259 Nagatsuta, Midori, Yokohama, Kanagawa, 226-8503, Japan \\ 
              \email{yshikano@ims.ac.jp}          
}

\date{Received: date / Accepted: date}

\maketitle

\begin{abstract}
Phase-squeezed light can enhance the precision of optical phase estimation. The larger the photon numbers are and the stronger the squeezing is, the better the precision will be. We propose an experimental scheme for generating phase-squeezed light pulses with large coherent amplitudes. In our scheme, one arm of a single-photon Mach--Zehnder interferometer interacts with coherent light via a non-linear optical Kerr medium to generate a coherent superposition state. Post-selecting the single photon by properly tuning a variable beam splitter in the interferometer yields a phase-squeezed output. 
\end{abstract}

\section{Introduction}
Optical phase estimation has many practical applications, such as metrology \cite{Giovannetti2011} laser interferometer gravitational wave detectors ~\cite{GEO,LIGO,Grav}, optical communications \cite{Bachor2004,Slavik2010,Chen2012}, quantum communications \cite{Loock2006,Loock2008}, and quantum computation \cite{Nemoto2004,Munro2005,Kok2007,Spiller2006}. Generally, the precision of phase estimation is limited by the standard quantum limit (SQL) \cite{Giovannetti2011,Braginsky1992}. To enhance the precision of optical phase estimation, two quantum-optics methods have been proposed: entangling of photons into a ``N00N'' state \cite{Giovannetti2011,Afek2010,Israel2012,Resch2007,Xiang2011,Ono2008,Dorner2009} and phase-squeezed light \cite{Bachor2004,Lvovsky2015,Gerry2005}. In a \NOON state, the entangled photon state is written as $(\ket{N}\ket{0}-\ket{0}\ket{N})/\sqrt{2}$, where $\ket{0}$ is the vacuum state and $\ket{N}$ is the number state with $N$ photons. This can improve the statistical scaling of the error from that of the SQL, $N^{-1/2}$, to that of the Heisenberg limit, $N^{-1}$, for quantum metrology \cite{Giovannetti2011,Resch2007,Xiang2011}. However, since post-selection is required to create the \NOON state \cite{Resch2007,Xiang2011}, the measurement is probabilistic. Moreover, it is technically difficult to implement the \NOON state with a large photon number \cite{Afek2010,D'Angelo2001,Mitchell2004,Kim2009}. The \NOON state with a photon number of $N=5$ has previously been experimentally achieved by the mixing of squeezed vacuum and coherent light \cite{Afek2010}. However, the generation of \NOON states with larger photon numbers is limited by photon losses.\par

The other approach, phase squeezing of light, suppresses the phase fluctuations in the coherent light to sub-SQL levels at the cost of amplitude fluctuations \cite{Bachor2004,Lvovsky2015,Gerry2005}. It has been noted that quantum metrology can be implemented by using phase-squeezed light \cite{Yonezawa2012}. Moreover, phase-squeezed light can be directly applied to coherent optical communications \cite{Bachor2004}. One application is in quantum repeaters using entangled atoms prepared by entanglement generation by communication \cite{Praxmeyer2010} to improve both the success probability of the entanglement generation and the fidelity between an ideal atomic Bell state and the generated one. Another is in quantum computation using atomic entanglement generation by communication \cite{Matsuoka2013} to suppress the error probability in the entanglement generation. To improve the precision with which these applications can be implemented, it is important to generate strongly phase-squeezed light pulses with large coherent amplitudes, i.e., the high peak power. \par
There are several theoretical proposals for generating phase-squeezed light \cite{Lvovsky2015,Gerry2005,Ralph1995,Levien1991}. As an example, the phase-squeezed state can be mathematically treated as a ``displaced'' squeezed vacuum state \cite{Lvovsky2015,Gerry2005,Yuen1976}. When such a displacement operation can be performed on a strongly squeezed vacuum state, strongly phase-squeezed light with large coherent amplitudes may be achieved. The displacement operation is implemented by the mixing of two modes of light using a beam splitter \cite{Paris1996} and the noiseless amplification of a squeezed vacuum \cite{Ralph1995}. While a squeezed vacuum with a squeezing of $12.7$ dB, generated with a $1064$ nm laser, has been observed \cite{Eberle2010}, a large displacement operation has not been reported. This may be because photon loss degrades the squeezing effect. As another example, phase squeezing using subharmonic generation has been proposed \cite{Levien1991}. An early experimental example is the observation of $0.7$-dB phase-squeezed light (continuous wave) at 1064 nm using a monolithic standing-wave lithium-niobate optical parametric amplifier with a mean photon number per sec of $\bar{n}=2.9\times 10^{6} \ {\rm s}^{-1}$ ($0.59$ pW) \cite{Breitenbach1997}. Recent experimental examples of using subharmonic generation include the observation of $3.2$-dB phase-squeezed light (continuous wave) at $860$ nm \ \cite{Yonezawa2012} and at $1064$ nm \ \cite{Xie2007} using a periodically poled ${\rm KTiOPO}_{4}$  with mean photon numbers per sec of $\bar{n}=1.0\times 10^{6} \ {\rm s}^{-1}$ ($0.24$ pW) and $\bar{n}=8.6\times 10^{14}\  {\rm s}^{-1}$ ($0.16$ mW), respectively. However, the generation of phase-squeezed light pulses with large coherent amplitudes has not yet been reported experimentally. \par
In this paper, we propose a scheme to perform phase squeezing on coherent light pulses with large coherent amplitudes. In our proposed scheme, approximate phase-squeezed state is generated by the quantum interference between two slightly phase-shifted coherent states followed by post-selection of the destructive interference. The phase shift is induced by the weak cross-Kerr modulation (XPM) with single photon. 
The proposed set up was considered by Feizpour et al. \cite{Feizpour2011}  for the weak value amplification \cite{Aharonov1988} of the single photon non-linearity on a coherent state. By careful analysis on the post-selection process, we found that quantum interference alters the coherent state to a phase-squeezed state. To verify the phase-squeezing effect, we calculated the maximum fidelity between the generated state and an arbitrary squeezed coherent state. We found that a $2.08$-dB phase-squeezed state can be generated with a fidelity of $F=0.99$ with a mean photon number of $\left| \alpha \right|^{2}=3.0\times 10^{6}$ and the full-width at half-maximum of approximately $0.6$ ps, i.e., mean photon numbers per sec $\bar{n}=5 \times 10^{18} \ {\rm s}^{-1}$ ($1.24$ W) under experimentally feasible values for inducing XPM, previously obtained in Ref. \cite{Matsuda2009}.

\section{Squeezing effect for coherent light}
\begin{figure}[t]
\begin{center}
\includegraphics[width=6cm]{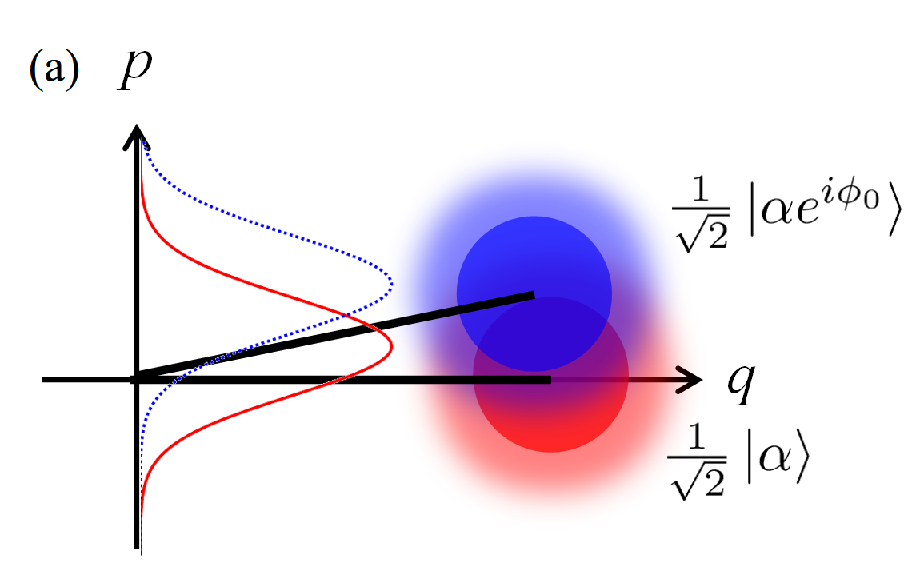} \\
\centering\includegraphics[width=8cm]{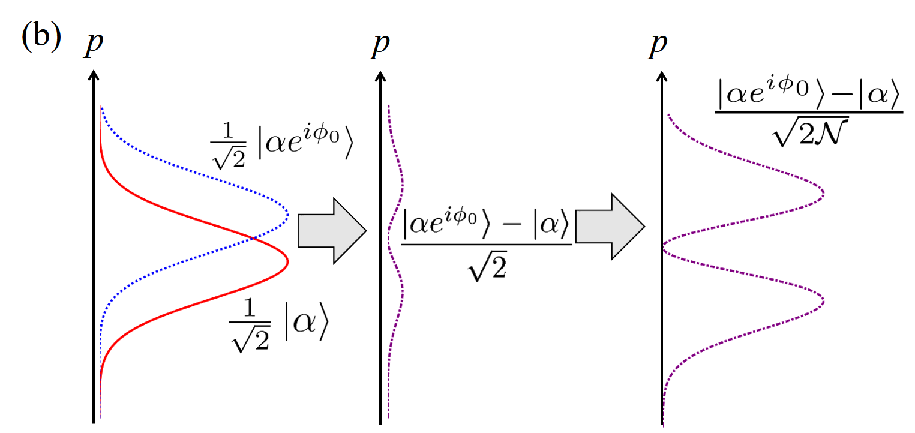} \\
\includegraphics[width=8cm]{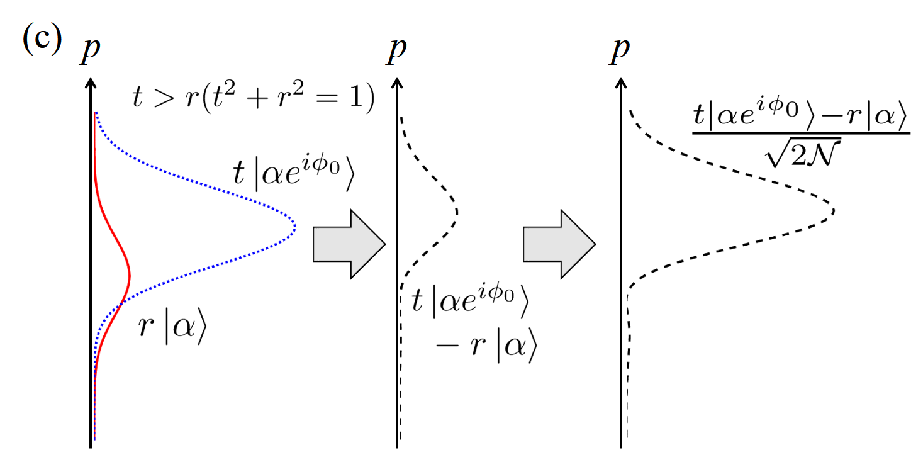}
\end{center}
\caption{\label{fig1}{\bf a} Phase-space representation and images of $p$-quadrature probability density distributions for each component of the superposition state in Eq. (\ref{CSSeq}). {\bf b} The $p$-quadrature probability density distributions of the components ({\it left}), without normalization of Eq. (\ref{CSSeq}) ({\it middle}) and with normalization of Eq. (\ref{CSSeq}) ({\it right}) for $t=r=1/\sqrt{2}$, and {\bf c} for $t \neq r \neq 1/\sqrt{2}$}
\end{figure}

Let us briefly explain how to generate the phase-squeezed state of light with large coherent amplitudes. The form of superposition states, i.e., quantum interference, in phase space can be changed according to the probability amplitudes and the relative phase between the components of the superposition. An approximate phase-squeezed state can be generated by controlling the probability amplitudes between the phase-shifted coherent state $\ket{\alpha e^{i \phi_{0}}}$ and the non-phase-shifted coherent state $\ket{\alpha}$ with a coherent amplitude $\alpha$, which are the components of the superposition state:
\begin{equation}
\ket{\psi_{{\rm CSS}}}=\frac{1}{\sqrt{2{\cal N}}}(t \ket{\alpha e^{i \phi_{0}}}-r\ket{\alpha}), \label{CSSeq}
\end{equation}
where $t$ and $r$ are assumed to be real numbers with $t^{2}+r^{2}=1$ and the normalized factor ${\cal N}$ is:
\begin{equation}
{\cal N} = \frac{1}{2} \left| t^{2}+r^{2}-tr (\braket{\alpha e^{i \phi_{0}}  | \alpha}+\braket{\alpha | \alpha e^{i \phi_{0}}}) \right|. \label{suceq}
\end{equation}
Figure \ref{fig1}a shows the phase-space diagram of the superposition state and images of the probability density distributions \cite{Barnett1997} $|\braket{p | \alpha }|^{2}$ and $|\braket{p | \alpha e^{i \phi_{0}}}|^{2}$ of the components of the superposition in Eq. (\ref{CSSeq}), which represent the outcomes of the quadrature measurement of the post-selected state projected onto the $p$-axis and correspond to phase measurement. Since the width of the $p$-quadrature distribution is determined by phase-fluctuations in the measured quantum states, the phase-squeezed effect can be verified by comparing the width of the distributions between the coherent state and Eq. (\ref{CSSeq}). For the superposition state of Eq. (\ref{CSSeq}), the form of the distribution $|\braket{p | \psi_{\rm CSS} }|^{2}$ depends on the probability of $|\braket{p | \alpha }|^{2}$ subtracted from $|\braket{p | \alpha e^{i \phi_{0}}}|^{2}$ and normalized by ${\cal N}$. As a special case where $t=r=1/\sqrt{2}$, i.e., an odd coherent state on $\phi_0 = \pi$, since the measurement probability does not give a negative value, two peaks appear in $|\braket{p | \psi_{\rm CSS} }|^{2}$, as shown in Fig. \ref{fig1}b. On the other hand, for $t \neq r \neq 1/\sqrt{2}$ and $t>r$, an asymmetric distribution appears, as shown in Fig. \ref{fig1}c, since the probability amplitudes of Eq. (\ref{CSSeq}) are asymmetrical. In this case, the edge of the distribution $|\braket{p | \alpha e^{i \phi_{0}}}|^{2}$ is subtracted by another distribution $|\braket{p | \alpha }|^{2}$ and almost no other peak with a negative value appears. Such a superposition state can be regarded as an approximate phase-squeezed state, since the width of the $p$-quadrature distribution is narrower than that of the coherent state. \par

\begin{figure}[t]
\centering\includegraphics[width=8cm]{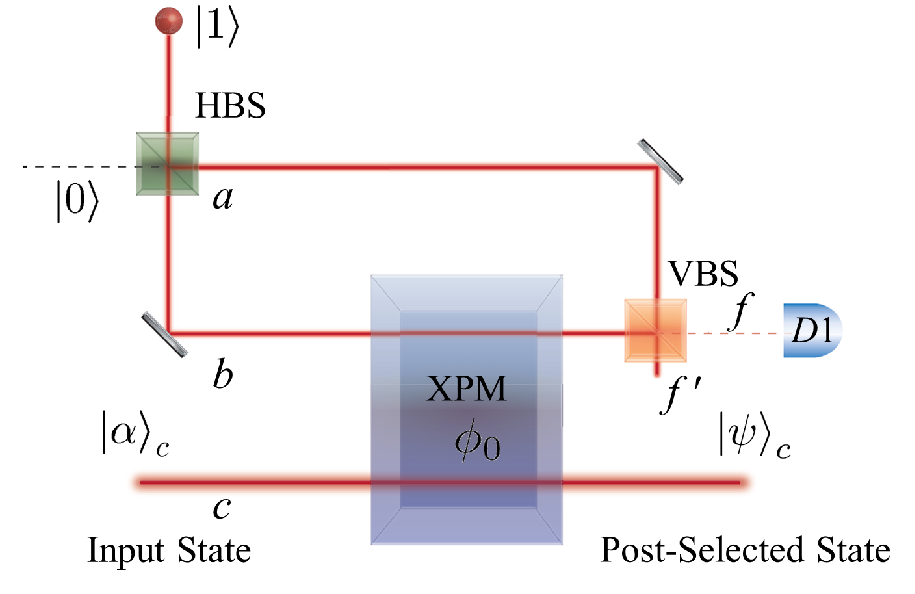}
\caption{\label{fig2} Our proposed setup. The post-selected state $\ket{\psi}_{c}$ can be regarded as the phase-squeezed state resulting from event selection at detector $D1$. It is noted that this proposed setup is also used in Ref. \cite{Feizpour2011} to amplify the single-photon non-linearity (color figure online)}
\end{figure}

Let us depict our proposed setup to implement such a squeezer. Our proposed setup, shown in Fig. \ref{fig2}, has a single-photon Mach--Zehnder interferometer with arms $a$ and $b$  and uses a non-linear optical Kerr medium to induce XPM on coherent light in arm $c$. In the single-photon Mach--Zehnder interferometer, the inputted single photon passes through a half beam splitter (HBS) and is divided into two arms, $a$ and $b$. The resulting state is written as $\ket{int}_{ab}\equiv (\ket{0}_{a}\ket{1}_{b}-\ket{1}_{a}\ket{0}_{b})/\sqrt{2}$, where $\ket{1}_{a}$ and $\ket{1}_{b}$ are single-photon states and $\ket{0}_{a}$ and $\ket{0}_{b}$ are vacuum states in arms $a$ and $b$, respectively. Then, the single photon in the two arms passes through a variable beam splitter (VBS) with transmissivity $t$ and reflectivity $r$ with $t^{2}+r^{2}=1$.  When the photon is detected by a detector $D1$ at the output port $f$, the quantum state of the photon is post-selected to be in  $\ket{f}_{ab}=t \ket{0}_{a}\ket{1}_{b}+r \ket{1}_{a}\ket{0}_{b}$.
Furthermore, the effect of the non-linear optical Kerr medium that is placed between arms $b$ and $c$ is represented by a unitary operator $\hat{U}={\rm exp} (i\phi_{0}\hat{n}_{b}\hat{n}_{c})$, where $\phi_{0}\ll 1$ is the phase shift angle caused by the XPM, and $\hat{n}_{b}$ and $\hat{n}_{c}$ are the photon number operators in arms $b$ and $c$, respectively. It is assumed that the input state of arm $c$ is a coherent photon state $\ket{\alpha}_{c}$. Using the non-linear optical Kerr medium, the initial total state $\ket{int}_{ab}\ket{\alpha}_{c}$ is transformed as: 
\begin{eqnarray} 
\ket{\Psi} &=& \hat{U}\ket{int}_{ab}\ket{{\alpha}}_{c} = \frac{1}{\sqrt{2}}(\ket{0}_{a}\ket{1}_{b}\ket{\alpha e^{i \phi_{0}}}_{c}-\ket{1}_{a}\ket{0}_{b}\ket{\alpha}_{c}).
\end{eqnarray}
When the single photon is detected at the photon detector $D1$ at port $f$, the total state $\ket{\Psi}$ is post-selected to $\ket{f}_{ab}$. The post-selected state $\ket{\psi}_{c}$ and the success probability $P_{{\rm suc}}$ of the post-selection $ { }_{ab}\braket{f | \Psi}$ are the same as Eqs. (\ref{CSSeq}) and (\ref{suceq}), respectively. \par

As alluded before, we discuss whether this post-selected state $\ket{\psi}_{c}$ can be regarded as the phase-squeezed state. Since the probability density distribution of the post-selected state $\ket{\psi}_{c}$ is the same to Fig. \ref{fig1}c, this state can be approximated as a squeezed coherent state $\ket{\xi e^{i2\theta}, \gamma e^{i\theta}}$ with the squeezing parameter $\xi =x e^{i\varphi} $ and the phase angle $\theta$. To verify this quantitatively, we evaluate the maximum fidelity $F= \left| \braket{\xi e^{i2\theta}, \gamma e^{i\theta}| \psi}_{c} \right|$ between the post-selected state $\ket{\psi}_{c}$ and the ideal squeezed coherent state $\ket{\xi e^{i2\theta}, \gamma e^{i\theta}}$ for $1/\sqrt{2}<t<1$ and evaluate two parameters, the squeezing parameter $\xi =x e^{i\varphi}$ and the phase angle $\theta$ for the maximized fidelity $F$. It is noted that the squeezing parameter $\xi =x e^{i\varphi}$ is characterized by the single-mode squeezing operator ~\cite{Barnett1997} $\hat{S}(\xi)\equiv e^{\frac{\xi}{2}(\hat{a}^{2}-(\hat{a}^{\dagger})^{2})}$ with the amplitude $x$ and phase $\varphi$. Using the squeezing operator and a coherent state  $\ket{\gamma}$ with an amplitude $\gamma$, the squeezed coherent state can be described as $\hat{S}(\xi)\ket{\gamma}=\ket{\xi, \gamma}$. Moreover, the phase angle $\theta$ is characterized by the phase-shifted squeezed coherent state $\hat{R}(\theta)\ket{\xi, \gamma}=\ket{\xi e^{i2\theta}, \gamma e^{i\theta}}$, where $\hat{R}(\theta)=e^{i\theta \hat{n}}$. Here, we assume that the mean photon numbers $\bar{n}$ of the coherent state $\ket{\alpha}$ and the squeezed coherent state $\ket{\xi e^{i2\theta}, \gamma e^{i\theta}}$ are unchanged since the XPM does not affect the photon number. Note that the coherent amplitude $\gamma$ of phase-squeezed state is different from the coherent state one $\alpha$ as $\gamma=\sqrt{\{ |\alpha|^{2}-{\rm sinh} r\} / \{ e^{-2r} {\rm cos}(\pi/2)+e^{2r} {\rm sin}(\pi/2) \}}$ under the same photon numbers. Furthermore, we assume $\alpha=10^{5/2}$ and $\phi_{0}=2\pi \times 10^{-5}$, which are the same values used in Ref. \cite{Feizpour2011}. \par
\begin{figure}[t]
\centering\includegraphics[width=6cm]{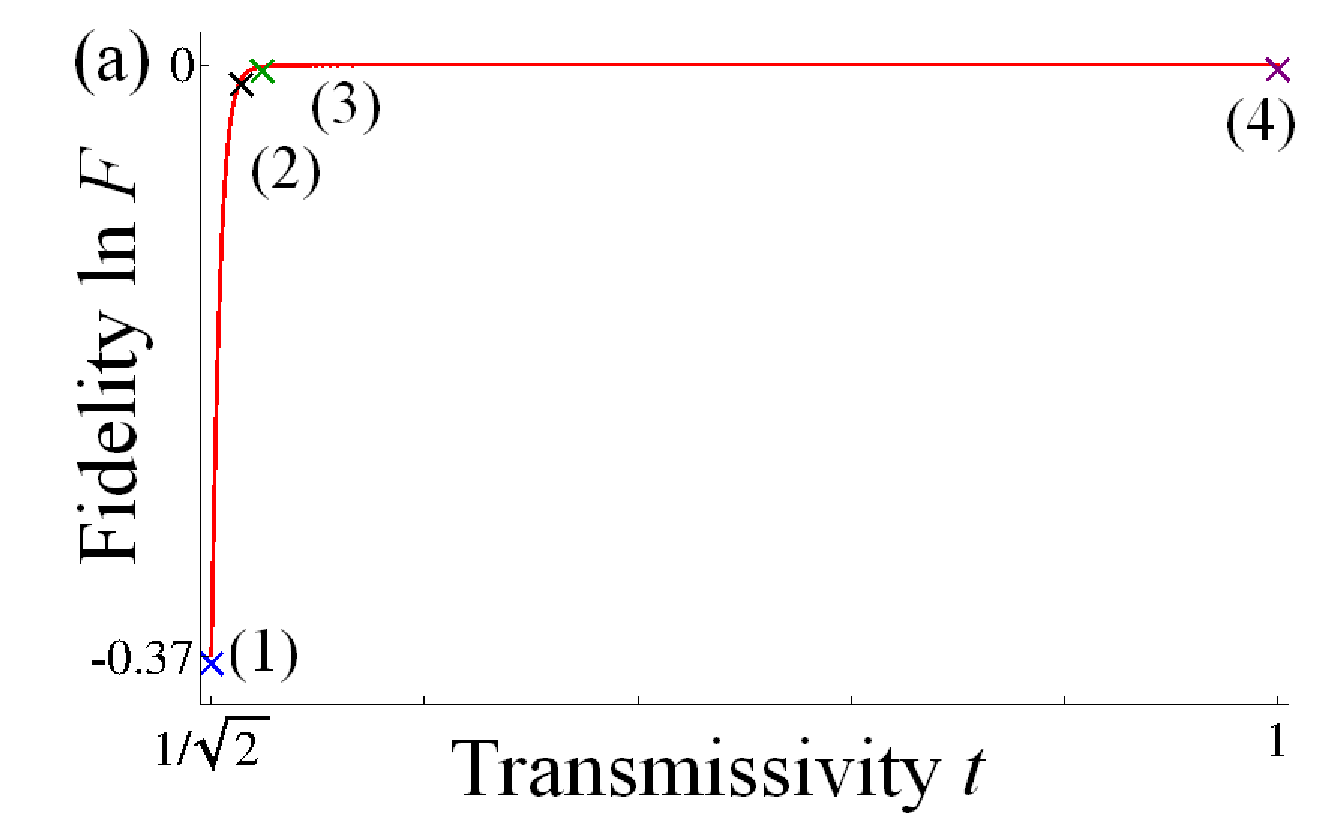} 
\centering\includegraphics[width=6cm]{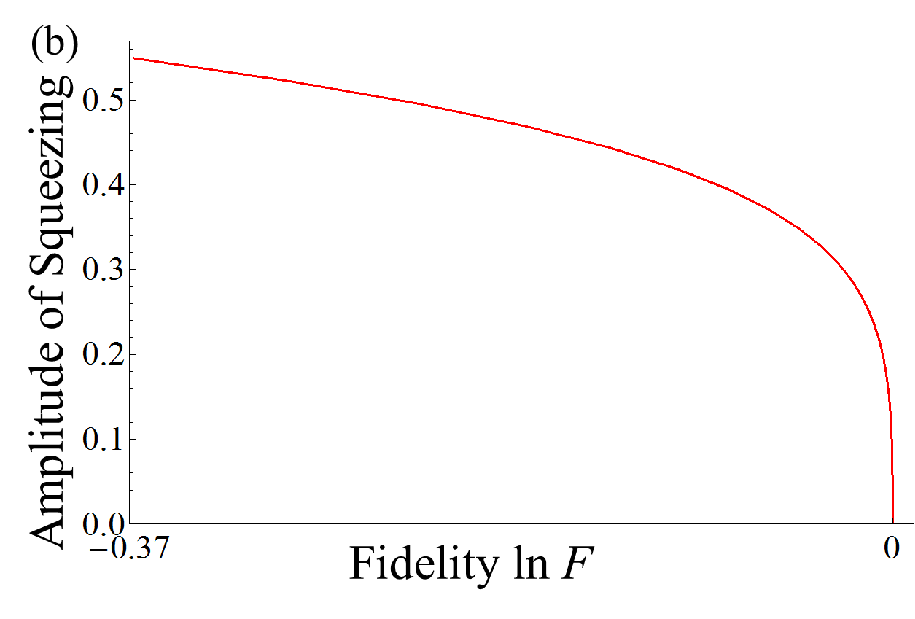} \\
\centering\includegraphics[width=6cm]{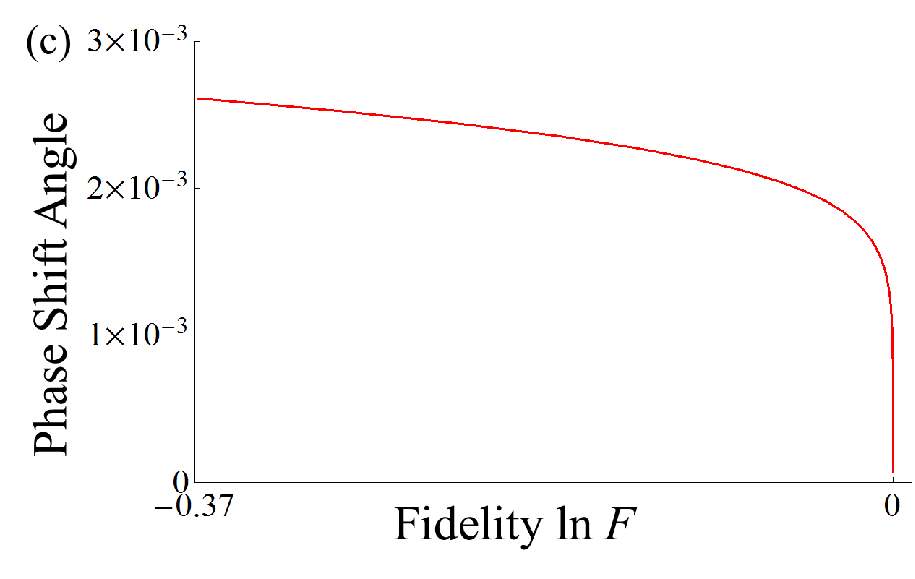} 
\centering\includegraphics[width=6cm]{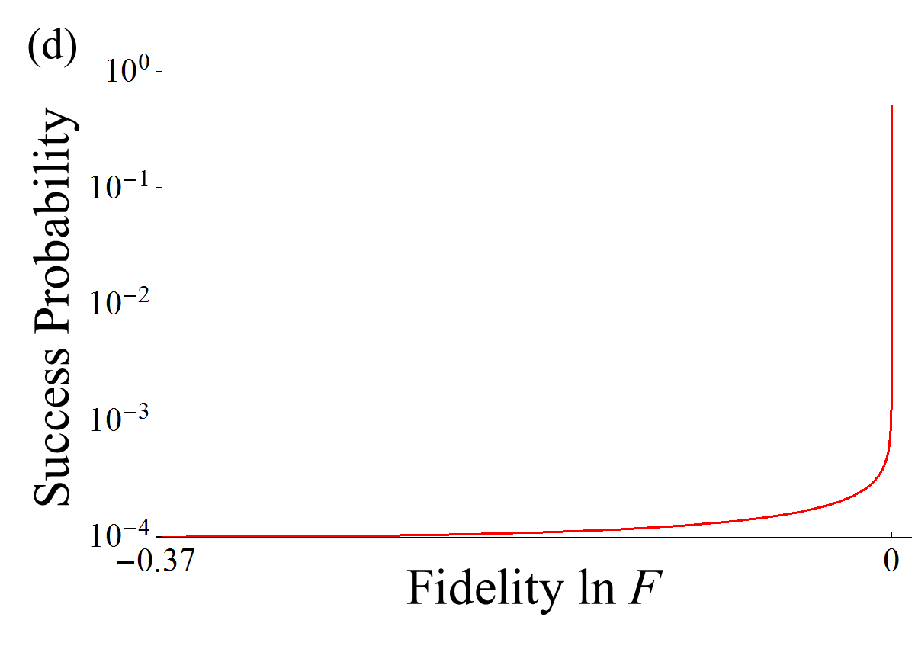} \\
\caption{\label{fig3}{\bf a} Numerical estimation of the maximum fidelity $F= \left| \braket{\xi e^{i2\theta}, \gamma e^{i\theta}| \psi}_{c} \right|$ as a function of the transmissivity $t$, ranging from $t=1/\sqrt{2}$ to $t= 1$. {\bf b} The estimated amplitude of the squeezing parameter of the post-selected state $x_{\rm est}$. {\bf c} The estimated phase $\theta_{\rm est}$. {\bf d} The success probability of the post-selection, $P_{\rm suc}$, as a function of the fidelity $F$. We note that the optimal value of the phase of the squeezing parameter to maximize the fidelity is $\varphi =\pi$. The points (1)--(4) in {\bf a} correspond to $t=1/\sqrt{2}$, $t=0.717$ ($F=0.99$), $t=0.724$ ($F=0.999$), and $t=1$, respectively}
\end{figure}

We numerically evaluated the maximized fidelity $F$ for various values of the transmissivity $t$; the results are plotted in Fig. \ref{fig3}a. The estimated parameters $\xi_{{\rm est}}=x_{{\rm est}} e^{i\varphi_{{\rm est}} }$ and $\theta_{{\rm est}}$ depend on the transmissivity $t$, and since the fidelity monotonically increases with transmissivity, these parameters can also be written as functions of the fidelity $F$. These parameters are plotted in Fig. \ref{fig3}b, c, respectively. Note that Fig. \ref{fig3}b shows the limitation of the obtainable amplitude of the squeezing parameter for $\alpha=10^{5/2}$ and $\phi_{0}=2\pi \times 10^{-5}$. The estimated parameter $\varphi_{{\rm est}}$ should always be $\pi$ to obtain the maximum fidelity $F$. This means that all estimated squeezed states are phase-squeezed states. Therefore, the post-selected state  $\ket{\psi}$ can be regarded as the phase-squeezed state for the high-fidelity cases. The success probability of the post-selection, $P_{{\rm suc}}$, for the fidelity $F$ is shown in Fig. \ref{fig3}d. The estimated parameters of the representative cases (1)--(4) in Fig. \ref{fig3}a are summarized in Table \ref{table1}. In case (1), since the post-selected state is equivalent to the odd-coherent-like state for $t=r=1/\sqrt{2}$, the fidelity, $F=0.69$, is low. In case (4), when the post-selection succeeds for a transmissivity of $t=1$, the single photon is transmitted in only arm $b$ of the Mach--Zehnder interferometer. Therefore, the post-selected state is just the phase-shifted coherent state $\ket{\alpha e^{i\phi_{0}}}_{c}$. These cases cannot be regarded as the phase-squeezed state. On the other hand, effective squeezing is achieved in both cases (2) and (3), where high fidelities are obtained $F=0.99$ for $t=0.717$ (2) and $F=0.999$ for $t=0.724$ (3), respectively. Therefore, these post-selected states can be regarded as quasi-phase-squeezed states. Herein, we refer to the post-selected state as the phase-squeezed state when the fidelity $F\geq 0.99$ and $x_{{\rm est}}\geq 0.01$ (approximately $8.0 \times 10^{-2}$ dB). \par

\begin{table}[t]
\centering\caption{\label{table1} Estimated parameters for maximizing the fidelity \protect $F$ for a given transmissivity \protect $t$, also given are the phase \protect $\theta_{{\rm est}}$, the amplitude of the squeezing parameter \protect $x_{{\rm est}}$, and the success probability of the post-selection, \protect $P_{{\rm suc}}$}
\begin{tabular}{cccccccc}
\hline
\textrm{\protect $t$}& \textrm{\protect $F$}& \textrm{\protect $x_{\rm est}$}& \textrm{\protect $\theta_{\rm est}$}& \multicolumn{1}{c}{\protect $P_{\rm suc}$} \\ 
\hline
1/\protect $\sqrt{2}$ & 0.69 & 0.55 & \protect $2.61\times 10^{-3}$ & \protect $9.87\times 10^{-5}$ \\
0.717 & 0.99 & 0.24 (2.08 dB) & \protect $1.60 \times 10^{-3}$ & \protect $2.95\times 10^{-4}$ \\
0.724 & 0.999 & 0.13 (1.13 dB) & \protect $1.15\times 10^{-3}$ & \protect $6.75\times 10^{-4}$ \\
1 & 1 & 0 & \protect $2\pi \times 10^{-5}$ & 0.5 \\ \hline
\end{tabular}
\end{table}

\begin{figure}[t]
\centering\includegraphics[width=6cm]{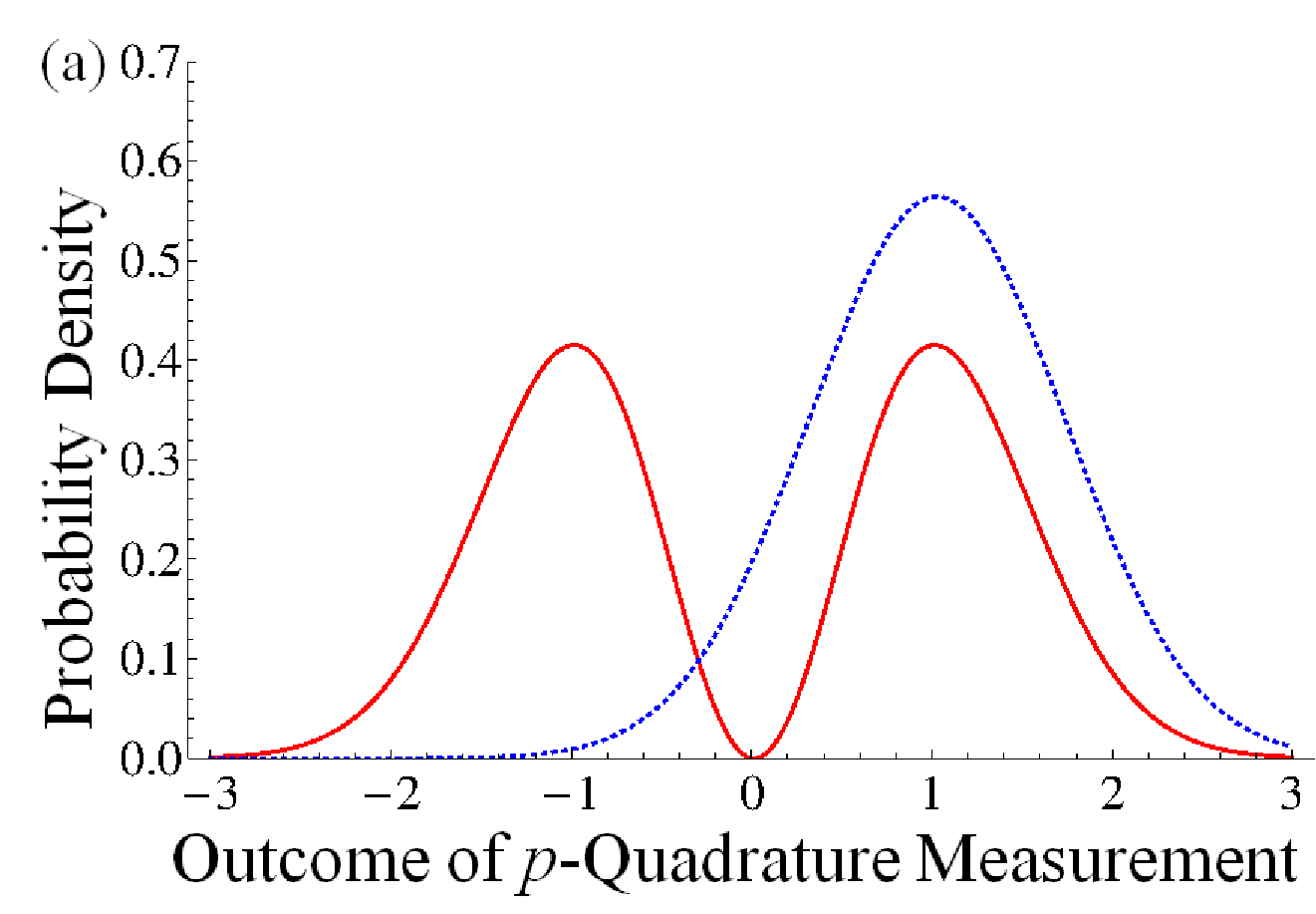} 
\centering\includegraphics[width=6cm]{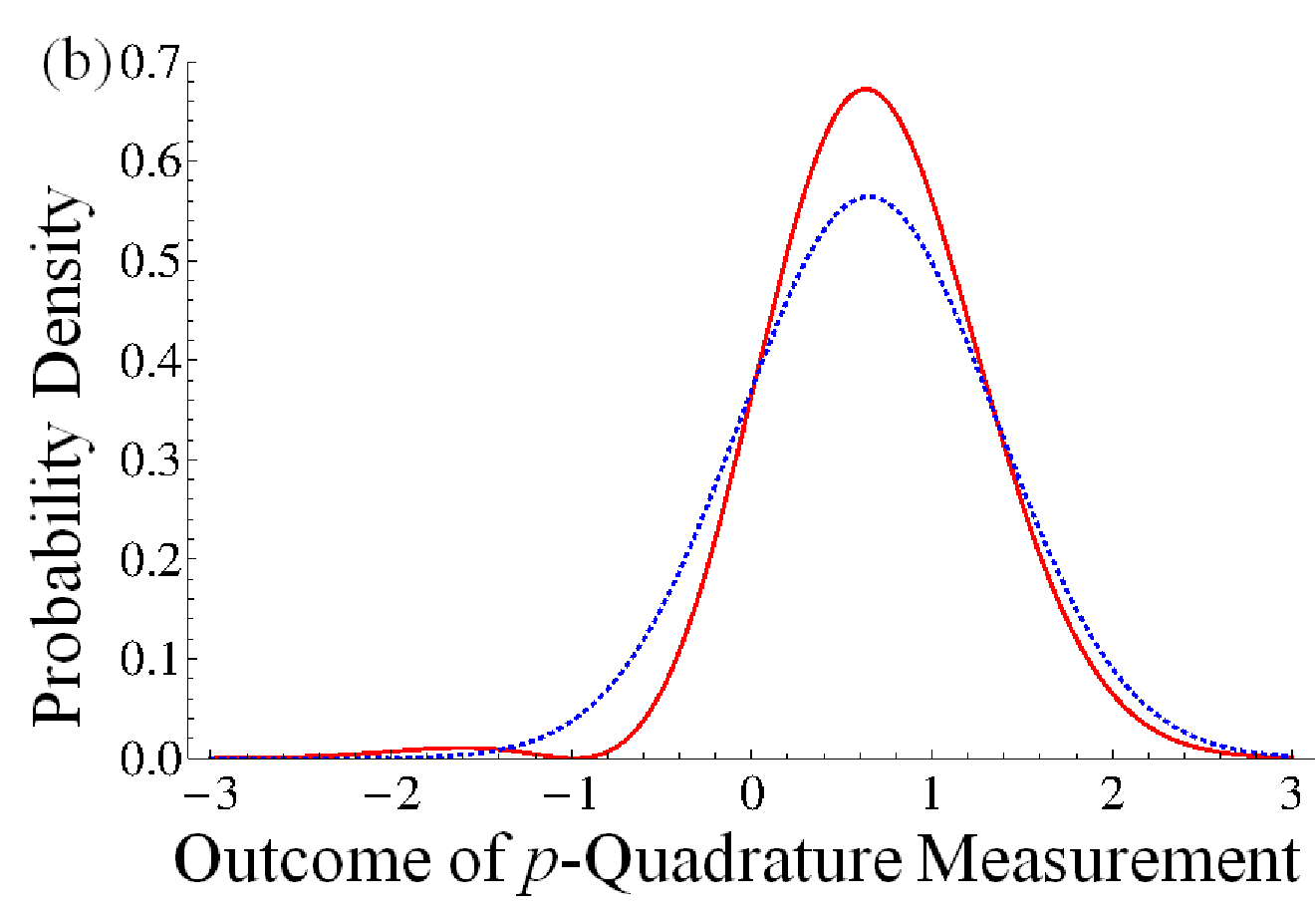} \\
\centering\includegraphics[width=6cm]{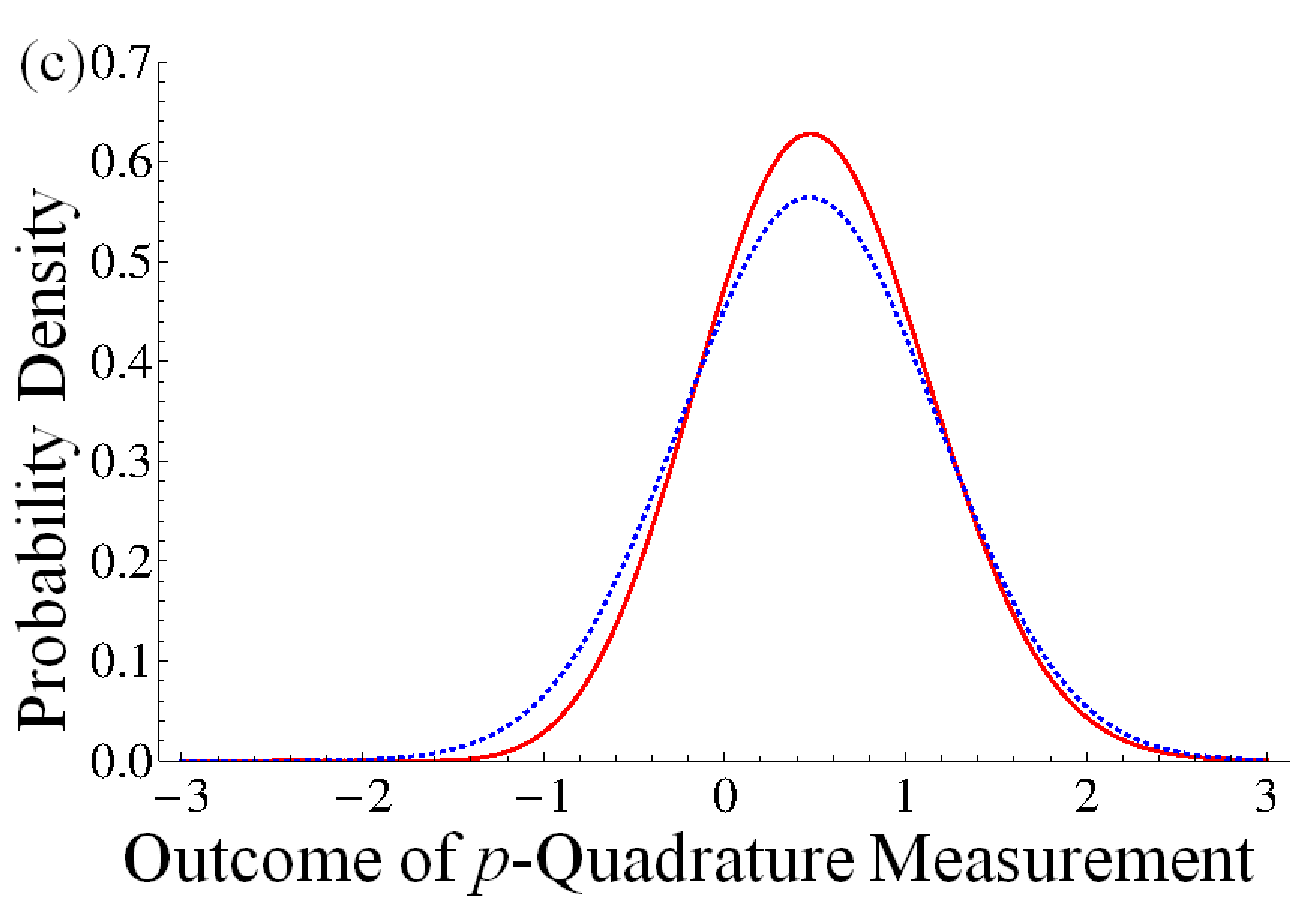} 
\centering\includegraphics[width=6cm]{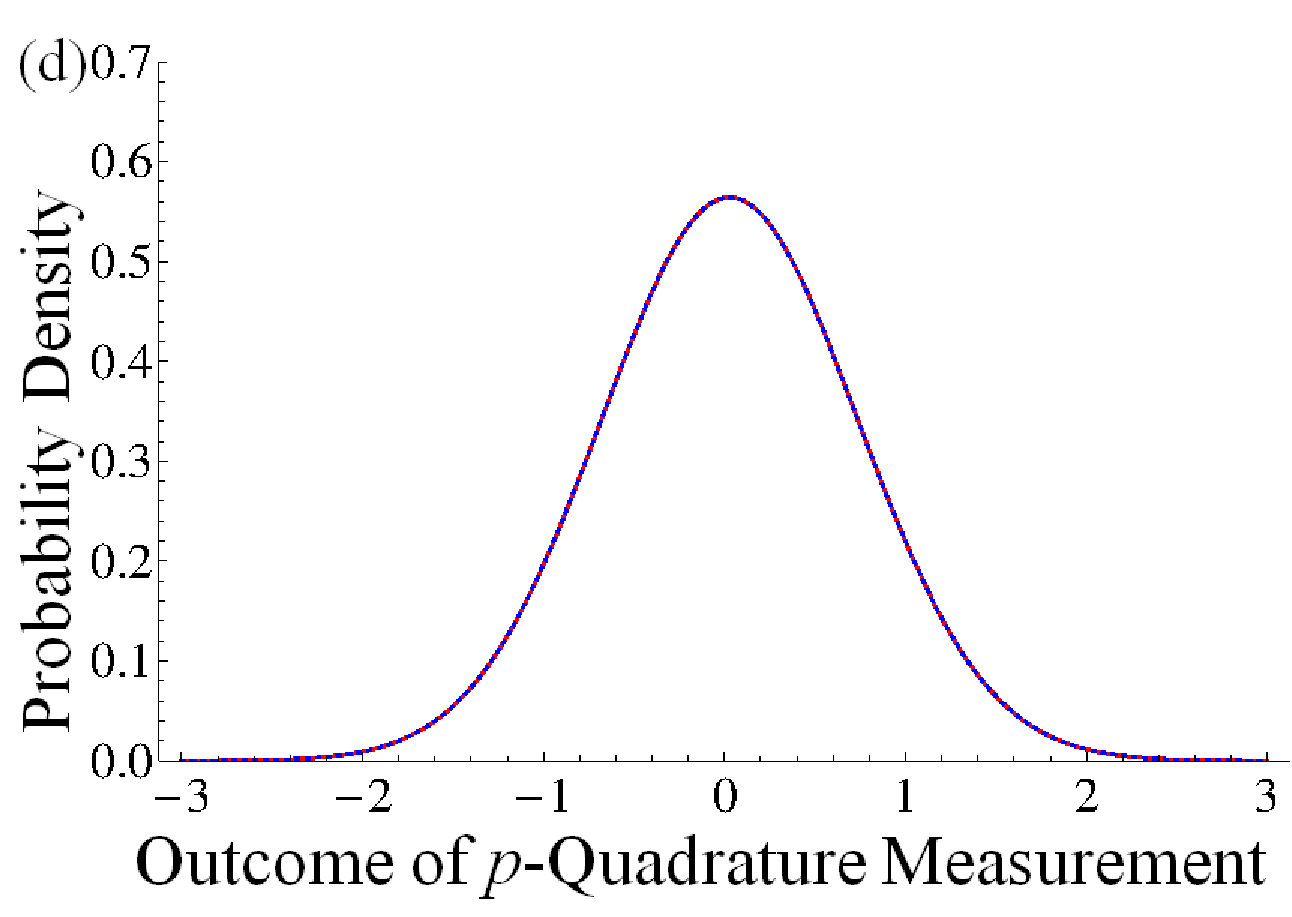} \\
\caption{\label{fig4} Probability density distributions of the post-selected state ({\it red solid line}) on the $p$-quadrature used to verify the squeezing effect. Here, panels {\bf a--d} correspond to cases (1) $t=1/\sqrt{2}$, (2) $t=0.717$ ($F=0.99$), (3) $t=0.724$ ($F=0.999$) and (4) $t=1$ in Fig. \ref{fig3}a, respectively. To compare these to the coherent state ({\it blue dotted line}), the peaks of the {\it blue dotted lines} are horizontally shifted to match those of the {\it red solid lines} (color figure online)}
\end{figure}

To verify the squeezing effect, we calculated probability density distributions $\left| \braket{p | \psi}_{c} \right|^{2}$ for the post-selected state using the specific parameters $\alpha=10^{5/2}$ and $\phi_{0}=2\pi \times 10^{-5}$, as shown in Fig. \ref{fig4}a--d, which correspond to cases (1)--(4) in Fig. \ref{fig3}a, respectively. For $t=1/\sqrt{2}$ as shown in Fig. \ref{fig4}a, two {\it squeezing-like} distributions, where the width of one peak is squeezed compared to the corresponding coherent state, are formed by the quantum interference in the overlap between $\ket{\alpha}_{c}$ and $\ket{\alpha e^{i\phi_{0}}}_{c}$ as a result of the post-selection. This state is, however, far from the squeezed states, because the probability distribution consists of two peaks with the same heights.  One of the peak can be post-selected by setting the interferometer asymmetric, i.e., $t \ne r \ne 1/\sqrt{2}$. For $t=1$, one peak of the probability distribution is completely suppressed, as shown in Fig. \ref{fig4}d; however, the resultant distribution shows no squeezing. Nevertheless, in the middle of these extremes, the quantum interference and the post-selection lead to an almost complete elimination of one peak of the probability distribution and the formation of a squeezed probability distribution compared to the Gaussian case, as shown in Fig. \ref{fig4}b, c. This is why our proposed scheme can be regarded as a phase squeezer.\par
\begin{figure}[t]
\centering\includegraphics[width=7cm]{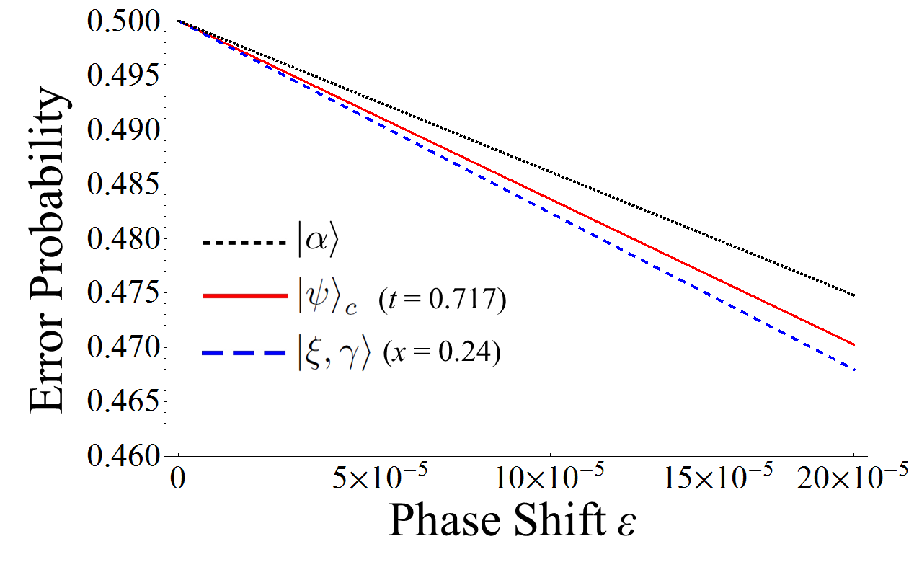} 
\caption{\label{fig5} The error probabilities of discrimination between non-phase-shifted and phase-shifted optical quantum states for small phase shifts $\epsilon$. Solid line is the post-selected states for $t=0.717$, which corresponds to Fig. \ref{fig4}b. {\it Dashed line} is the ideal phase-squeezed state $\ket{\xi, \gamma}$ for the amplitude of the squeezing $x=0.24$. Dotted line is the coherent states.}
\end{figure}

To quantitatively confirm the improvement of the phase sensitivity using the post-selected state, we numerically evaluated the error probability of discrimination of non-phase-shifted and non-phase-shifted optical quantum states for small phase shifts $\epsilon$; e.g., for the post-selected state, the overlap of the probability density distributions between $\ket{\psi}_{c}$ and $\ket{\psi e^{i \epsilon}}_{c}$, see Fig. \ref{fig5}. The solid line of Fig. \ref{fig5} is the post-selected states for the transmissivities of the VBS with $t=0.717$, which corresponds to the condition of Fig. \ref{fig4}b. The dashed line is the ideal phase-squeezed state $\ket{\xi, \gamma}$ for the amplitude of the squeezing $x=0.24$, which is the estimated parameter for maximizing the fidelity $F$, as shown in Table \ref{table1}. Here, the squeezing angle is $\varphi=\pi$ for the phase-squeezed state, i.e.,  $\xi=xe^{i \pi}$. The dotted line is the coherent states. Clearly, the error probability for the post-selected state is always smaller than coherent states. Thus, the post-selected state has higher sensitivity than coherent states.\par
\begin{figure}[t]
\centering\includegraphics[width=7cm]{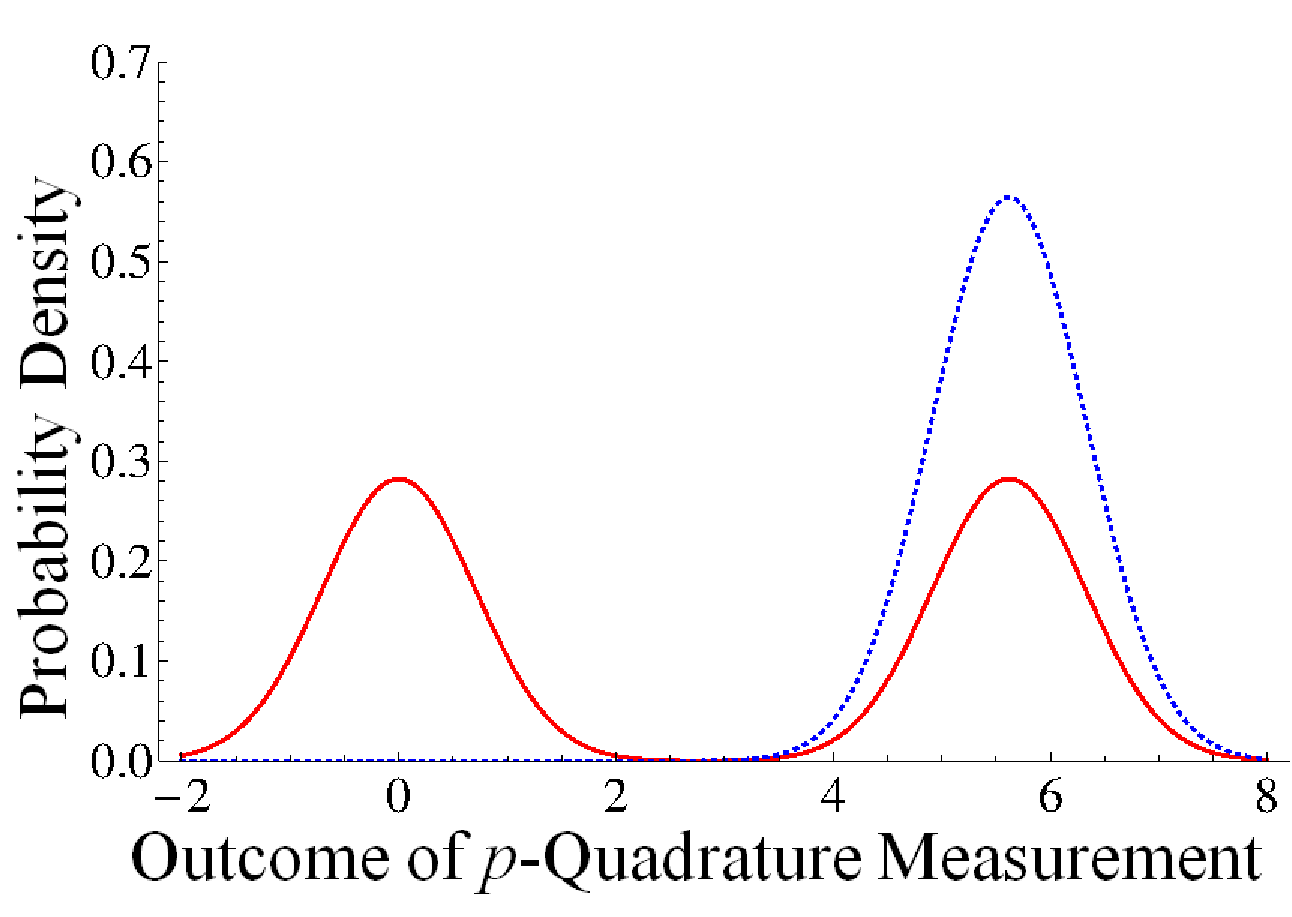} 
\caption{\label{fig6} Probability density distributions of the post-selected state ({\it red solid line}) and the coherent state ({\it blue dotted line}) used to verify the elimination of the squeezing effect on the $p$-quadrature in the case of $t=1/\sqrt{2}$ with $\left| \braket{\alpha | \alpha e^{i\phi_{0}}}_{c} \right| = 3.72 \times 10^{-4}$, in which a non-phase-shifted coherent state $\ket{\alpha}_{c}$ and a phase-shifted coherent state $\ket{\alpha e^{i\phi_{0}}}_{c}$ are far from overlapping (color figure online)}
\end{figure}

Furthermore, when the overlap $\braket{\alpha | \alpha e^{i\phi_{0}}}_{c}$ is very small, i.e., the distance between the states $\ket{\alpha}_{c}$ and $\ket{\alpha e^{i\phi_{0}}}_{c}$ is very large, the effect of the quantum interference between two well-separated peaks is minuscule. Therefore the squeezing effect does not occur, as shown in Fig. \ref{fig6}. We numerically confirm that using $\phi_{0} \gtrsim 0.01$ with $\alpha=10^{5/2}$ does not achieve effective squeezing for any transmissivity value. An indication of the optimum overlap $\braket{\alpha | \alpha e^{i\phi_{0}}}_{c}$ is $\alpha \phi_{0} \gtrsim 0.1$, since  $\braket{\alpha | \alpha e^{i\phi_{0}}}_{c}$ rapidly decreases for $\alpha \phi_{0} > 0.1$. Since the success probability is dependent on the overlap from Eq. (\ref{suceq}), it suggests that a higher success probability can be produced by a larger phase shift $\phi_{0}$ and coherent amplitude $\alpha$. For example, $\alpha \phi_{0} = 0.1$, $F=0.99$, and  $x_{\rm est}=0.24$ can be obtained with $P_{{\rm suc}}=7.08\times 10^{-3}$ by tuning the transmissivity to $t=0.753$. However, $\phi_{0}$ and $\alpha$ are restricted in experimental situations, see the next section. \par


We note that the back-action of the coherent light on the single-photon interferometer should be compensated. The interaction between the coherent state and the single-photon interferometer induces a relative phase shift between the arms of the interferometer, which may be represented $\ket{0}_{a}\ket{1}_{b}-\ket{1}_{a}\ket{0}_{b}\rightarrow e^{i \left| \alpha \right|^{2}\phi_{0}}\ket{0}_{a}\ket{1}_{b}-\ket{1}_{a}\ket{0}_{b}$. The squeezing effect can be achieved only when $\left| \alpha \right|^{2}\phi_{0}$ is near a multiple of $2\pi$. This is because the phase of the post-selected state also changes by the relative phase shift $e^{i\left| \alpha \right|^{2}\phi_{0}}$ in the interferometer. For example, for $\left| \alpha \right|^{2}\phi_{0}=\pi$ using $\alpha=10^{5/2}$ and $\phi_{0}=\pi \times 10^{-5}$, the post-selected state with $t=1/\sqrt{2}$ is approximately the coherent state. Therefore, it is necessary to use a phase shifter on the single-photon interferometer to be compensated when $\left| \alpha \right|^{2}\phi_{0}$ is not near a multiple of $2\pi$, to produce the phase-squeezing effect. \par

\section{Experimental feasibility}
Let us discuss the experimental feasibility of our scheme by considering previously an established method for inducing XPM with a single-photon-level non-linearity. In Ref. \cite{Matsuda2009}, the cross-Kerr-induced phase shift of $10^{-7}$ rad was measured in a photonic crystal fiber for coherent light pulses with the full-width at half-maximum of approximately $0.6$ ps and a mean photon number of $\left| \alpha \right|^{2}=3.0\times 10^{6}$, which corresponds to the peak power of $1.24$ W, at single-photon-level intensities by averaging over $3\times 10^{9}$ pulses at a repetition frequency of $1$ GHz at room temperature. Since a pulsed laser with a wavelength of $802$ nm is used, the photon loss of the homodyne measurement can be reduced using a Si photodetector \cite{Korde1987} that operates at around $800$ nm. We note that the total photon loss of the homodyne measurement is $\thicksim 0.07$ at $860$ nm \cite{Takeno2007}. Here, as mentioned above, we assume that the effect of photon losses on the generated light is negligible. In addition, photon loss in the Mach--Zehnder interferometer is also negligible because of the event selection.\par
For $\phi_{0}=10^{-7}$ and $F=0.99$, the amplitude of the squeezing of the post-selected state can be determined to be $x_{{\rm est}}=0.24$ ($2.08$ dB) with a phase shift angle of $\theta_{{\rm est}}=2.88\times 10^{-4}$ and a probability $P_{{\rm suc}}=2.19\times10^{-8}$. Under these conditions, the transmissivity should be tuned to $t=0.70719$. Similarly, for $\phi_{0}=10^{-7}$ and $F=0.999$, $x_{{\rm est}}=0.13$ ($1.13$ dB) is found with $\theta_{{\rm est}}=2.06\times 10^{-4}$ and $P_{{\rm suc}}=5.06\times10^{-8}$. In this case, the transmissivity should be tuned to $t=0.70725$. As mentioned in the previous section, the success probability can be improved by increasing $\phi_{0}$ and $\alpha$. However, $\phi_{0}$ is mostly decided by the material of the non-linear optical Kerr medium. Moreover, $\alpha$ is restricted by self-Kerr effect and sequential second-order non-linear effects, which become significant for large photon numbers. Therefore, to improve the success probability, a material that only induces a stronger XPM is required. Such stronger XPM may be achieved using electromagnetically induced transparency, since enhancing XPM in the short-pulse regime is experimentally observed \cite{Dmochowski2016}. \par

\section{Conclusion and discussions}
We proposed an experimental scheme for generating phase-squeezed light pulses through the post-selection of single photon coupled with coherent light pulses via a weak cross-Kerr non-linearity. To implement the post-selection of single photon, a Mach--Zehnder interferometer with a variable beam splitter (VBS) as an output was used. When one arm of the Mach--Zehnder interferometer interacted with coherent light via the weak cross-Kerr non-linearity, a superposition of the non-phase-shifted coherent state $\ket{\alpha}_{c}$ and the phase-shifted coherent state $\ket{\alpha e^{i\phi_{0}}}_{c}$  was generated. The post-selection of the single photon caused quantum interference between $\ket{\alpha}_{c}$ and $\ket{\alpha e^{i\phi_{0}}}_{c}$. When the transmissivity and the reflectivity of the VBS were properly set, effective squeezing could be obtained such that the output had high fidelity to the ideal phase-squeezed state. It should be noted that the squeezing results not directly from the Kerr non-linearity but from the post-selected superposition of the slightly sifted coherent states. \par
Our method can generate the phase-squeezed pulses, which have $10^{4}$ times more peak power than conventional methods and can be directly used to the measurement of the small phase shift. However, our proposed scheme is probabilistically operated and has a limit in the degree of squeezing for a single run. The repeated runs on our proposed scheme may enhance the amplitude of the squeezing, although the fidelity and successful probability may become lower than those obtained on the single run. \par
In this paper, we evaluated the fidelity \protect \linebreak $F= \left| \braket{\xi e^{i2\theta}, \gamma e^{i\theta}| \psi}_{c} \right|$ to confirm whether the post-selected state $\ket{\psi}_{c}$ could be regarded as the phase-squeezed state $\ket{\xi e^{i2\theta}, \gamma e^{i\theta}}$ with the angle of squeezing $\varphi =  \pi$. We confirmed that the post-selected state $\ket{\psi}_{c}$ was surely approximated as the phase-squeezed state, since  the numerical evaluation of the fidelity in Fig. \ref{fig3}a showed that the fidelity was always maximized for the estimated angle of squeezing $\varphi_{{\rm est}}=\pi$. This fact cannot be confirmed by the discussion on the phase noise. Furthermore, the phase angle $\theta$ is important for the actual phase measurements to determine the best projection angle. We can obtain the highest power to discriminate the non-phase-shifted state $\ket{\psi}_{c}$ and the non-phase-shifted one $\ket{\psi e^{i \epsilon}}_{c}$ in the experiment by setting the projection angle to the the estimated phase angle $\theta_{{\rm est}}$. \par
A few further considerations remain with regard to our proposed scheme. First, to attain a more realistic model of the experimental situation, the squeezing effect with photon losses using a beam splitter model should be considered \cite{Leonhardt1997}. Notably, the photon loss and the self-Kerr induced anti-phase squeezing \cite{Matsuda2009,Imoto1985,Yamamoto1999,Kitagawa1986} for the post-selected state would weaken the squeezing effect. On the other hand, the photon loss for the single-photon interferometer would decrease the success rate of the post-selection. Moreover, the distortion effect, which changes the mode shape of the single photon by the XPM, should be considered, since the success probability of the post-selection will also be changed. Since the amplification of the XPM on a continuous wave probe field by the post-selection of the single photon is observed  \cite{Feizpour2015}, such distortion effect will not serious problem. Second, our proposal suggests that the weak-value amplification for the phase shift of coherent light through the post-selection of single photon \cite{Feizpour2011,Feizpour2015} should be reconsidered in view of the squeezing effect by the quantum interference. Our estimated phases are also amplified, as shown in Table \ref{table1}. As alluded before, this squeezing mechanism is based on the weak-value amplification of the single-photon non-linearity \cite{Feizpour2011}. While the advantage of the weak-value amplification may be caused by the squeezing effect of the quantum state ~\cite{Susa,Shikano,Pang,Yusuf}, the general relationship has not been found. Furthermore, the relationship between quantum interference and the weak values has already been discussed \cite{Tamate2009,Dressel2015}. Since the argument of the weak value is taken as the geometric phase, the relative phase may be characterized by the geometric phase of the single-photon interferometer. Finally, our scheme may be generalized to an arbitrary quadrature squeezer, since quantum interference can be controlled by changing the relative phase. Note that the effect of the phase noise is varied according to arbitrary quadrature-squeezed states. For example, the self-Kerr effect enhances the squeezing amplitude for amplitude squeezed states, since anti-phase squeezing occurred. Therefore, since the experimental feasibility is also varied according to the quadrature, the phase noise effect for arbitrary quadrature squeezing should be considered.

\section*{Acknowledgment}
The authors thank with Nobuyuki Matsuda, Hirokazu Kobayashi, Yoichi Aso, and Yu-Xiang Zhang for their useful discussions. F. M. and Y. S. thank Kyoko Kamo and Mayuko Kato for technical support on the figure illustrations. F. M. thanks the Institute for Molecular Science (IMS) for their hospitality, and is financially supported by IMS Joint Study. This work is also supported by the Center for the Promotion of Integrated Sciences (CPIS) of Sokendai. Y. S. was supported in part by Perimeter Institute for Theoretical Physics. Research at Perimeter Institute is supported by the Government of Canada through Industry Canada and by the Providence of Ontario through the Ministry of Economic Development \& Innovation.

\end{document}